\documentclass[a4paper,11pt]{article}

\usepackage{pos}
\usepackage{svg}
\usepackage{units}
\usepackage[capitalize]{cleveref}

\let\OLDthebibliography\thebibliography
\renewcommand\thebibliography[1]{
  \OLDthebibliography{#1}
  \setlength{\parskip}{0pt}
  \setlength{\itemsep}{0pt plus 0.3ex}
}

\usepackage{setspace}

\title{Combining IceCube Muon Tracks and Cascades to measure the Galactic Diffuse Neutrino Flux}

\ShortTitle{Combining IceCube Muon Tracks and Cascades to measure the Galactic Diffuse Neutrino Flux}

\author{The IceCube Collaboration \\{\normalsize \normalfont(a complete list of authors can be found at the end of the proceedings)}\\}

\emailAdd{jonas.hellrung@ruhr-uni-bochum.de}
\emailAdd{julia.tjus@ruhr-uni-bochum.de}
\emailAdd{wolfgang.rhode@tu-dortmund.de}

\abstract{
The diffuse Galactic neutrino flux is produced by cosmic rays interacting with the interstellar medium. The measurement of this flux can help to understand the distribution of cosmic rays in the Galaxy. The first observation of this neutrino flux was published in 2023 by the IceCube Collaboration. Here, plans for a new analysis combining different event topologies are presented. IceCube measures events in two main topologies. Tracks, originating in charged current $\nu_\mu$ interactions, provide a better angular resolution. In contrast, cascades, from most other possible interactions, provide a better energy resolution and are able to observe the Southern sky (and therefore the Galactic Center) despite the huge background of atmospheric muons. Combining both event topologies in one analysis exploits all these advantages. Sensitivities and model discrimination power of a combined measurement using a forward folding binned likelihood fit are discussed here. 

\vspace{4mm}

{\bfseries Corresponding authors:}
Jonas Hellrung$^{1,2*}$, 
Julia Becker Tjus$^{1,2,3}$, 
Wolfgang Rhode$^{2,4}$, 
\\
{$^{1}$ \itshape Theoretical Physics IV, Faculty for Physics and Astronomy, Ruhr University Bochum, Germany}\\
{$^{2}$ \itshape Ruhr Astroparticle and Plasma Physics Center (RAPP Center), Germany}\\
{$^{3}$ \itshape Department of Space, Earth and Environment, Chalmers University of Technology,  Sweden}\\
{$^{4}$ \itshape Department of Physics, TU Dortmund University, Germany}\\[4mm]
$^*$ Presenter
}

\ConferenceLogo{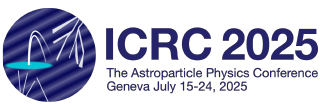}

\FullConference{39th International Cosmic Ray Conference (ICRC2025)\\
 15–24 July 2025\\
Geneva, Switzerland\\}

\begin{document}
\newcommand{\fermi}{Fermi-$\pi_0$}
\newcommand{\krafive}{KRA$_\gamma^5$}
\newcommand{\krafifty}{KRA$_\gamma^{50}$}

\maketitle

\section{Introduction}
In 2023, IceCube published the first observation of high-energy neutrinos from the Galactic plane \cite{GP_paper}. A part of this neutrino flux is expected to come from Galactic diffuse emission which arises when high-energy cosmic rays interact with the interstellar gas in the Milky Way. Additionally, unresolved neutrino sources in the Galaxy could also contribute to this neutrino flux. The measurement of the properties of this emission can help to understand the distribution of cosmic rays in our Galaxy. A better understanding of this distribution would be an important input to constrain cosmic ray models and to find the sources of cosmic rays in our Galaxy.

The first measurement of the Galactic neutrino flux used a selection of cascade-like events. These events can be seen from all directions but have a relatively large directional uncertainty ($\sim10^\circ$)\cite{GP_paper}. Since then, hints for a Galactic neutrino flux were also found in other event selections \cite{philipp_icrc,ESTES} and also by the ANTARES experiment \cite{antares_ridge}.

To better understand the Galactic neutrino flux, we want to use the approach established in \cite{combined_fit_icrc} and combine selections targeting cascade-like and track-like events to exploit their respective advantages. 

\section{Detector and Event Selections}
IceCube is a cubic kilometer neutrino detector at the South Pole. It consists of more than 5000 digital optical models which are embedded in a depth of \unit[1450]{m} to \unit[2450]{m} in the thick Antarctic ice shield and record light emission in the ice \cite{Icecube_Instrumentation}. This allows IceCube to measure the Cherenkov light emitted when high-energy charged particles move through the ice. Most of these particles are muons created in cosmic-ray air showers, but also secondary particles from neutrinos interacting in the ice or the surrounding rock are measured.

\subsection{Tracks}
One strategy is to search for long track-like events, that are produced by muons. But only a small fraction of muons is produced by charged current interactions of muon neutrinos and most of them originate in cosmic-ray air showers. As the muons can only travel up to a few kilometers before losing their energy, one can use the Earth as a shield and select only events which are reconstructed to be up-going. For this purpose a zenith cut of $\theta > 85^\circ$ is applied. Additionally, boosted decision trees are used to separate high-quality tracks from badly reconstructed or more spherical cascade-like events. The track sample used here is the same as described in \cite{combined_fit_icrc,philipp_icrc}. The sample has a high purity of >\unit[99.8]{\%} and a median angular resolution < $1^\circ$. The interaction of the neutrino does not need to happen inside of the detector, as long as the muon enters the detector at some poiade event selections, as visible in \cref{aeff}. But, as this sample only contains the northern hemisphere, and therefore only covers the Galactic plane for longitudes < -141$^\circ$ and > 27$^\circ$, it can not be used to observe the Galactic center and therefore a large fraction of the neutrino emission from the Galactic plane.

\subsection{Cascades}
\begin{figure}[t!]
\centering
\includesvg[width=0.6\linewidth]{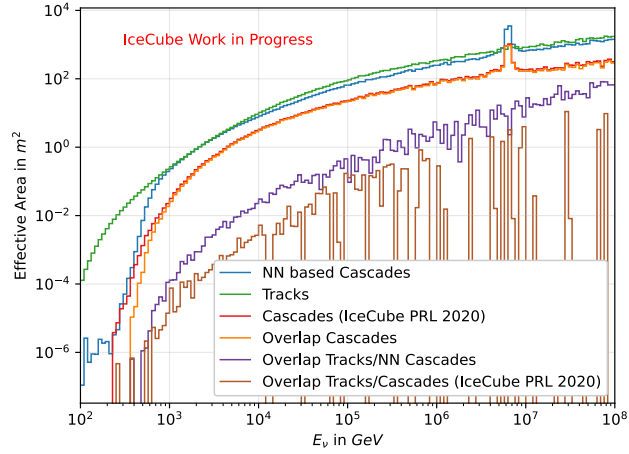}
\caption{Effective area for the used event selections. Green shows the track selection and Blue the NN based cascade selection. For comparison the cascade selection used in the previous combined fit is also shown in red. The other lines show the effective area of the overlap between the different selections.}\label{aeff}
\end{figure}

A second strategy is to look for cascade-like events. These events are created when the neutrino undergoes a neutral current interaction or if an electron or tau neutrino interacts via a charged current interaction. The secondary particles lose their energy in a shorter distance in this case which leads to a spherical signature. The selection used here is described in more detail in \cite{GP_paper}. It uses multiple neural networks to select neutrino events and reduce the atmospheric muon background. The energy and the direction of the event are also reconstructed with a convolutional-neural-network based approach \cite{CNNreco}. The advantage of this selection can be seen in \cref{aeff}. The effective area of this selection is almost as high as the one for the track selection and is about 3 times larger than the effective area of the cascade sample used in \cite{6_year_cascades} and \cite{combined_fit_icrc}. \cref{aeff} also shows that the overlap between the track sample and the cascade samples is quite small (<1\%). To combine the selections, the overlap is removed from the cascade selection as it mostly consists of starting tracks, which are better reconstructed in the track selection.

\section{Analysis Method}

\subsection{Forward-folding Likelihood Fit}
The analysis is performed with a forward-folding likelihood fit, where simulated neutrino events are weighted to calculate the expected number of events according to the model parameters. These events are then binned in reconstructed energy, zenith and right ascension. This is done separately for the two event selections. Then a Poisson likelihood is used to calculate the probability for the parameter values for the given data. We plan to use the effective likelihood \cite{SAY}, which takes the statistical uncertainties of the produced simulation into account, in the final analysis.\\
The computationally demanding task of calculating the likelihood is performed with an \textit{aesara}-based framework which was already used in previous analysis e.g. \cite{philipp_icrc,combined_fit_icrc,9.5yearsDiffuse}

\subsection{Modeling of Systematic Effects}
For the modeling of the detector response and optical properties of the ice at the South Pole, the SnowStorm method \cite{SnowStorm} is used. In this method, the parameters describing the ice and the detector are randomly sampled for each event during the simulation. The allowed ranges for the parameters were determined from calibration data. The simulation is split into two parts for each parameter and the gradient of the number of events per bin is calculated. This gradient can then be added to the expected number of events which is calculated from a baseline simulation where the parameters are fixed. This approach was developed and verified for \cite{combined_fit_icrc}. For the tests presented here, the systematic parameters are only applied to the track selection because of limited availability of simulations. In the final analysis, it will be included for both event selections in the same way.

\subsection{Background Modeling}
There are multiple backgrounds to the analysis. The first one is the isotropic astrophysical neutrino flux. It was discovered in 2013 \cite{IcecubeFirstAstroSignal} and characterized in multiple analysis since then. Recently there was evidence for deviation from a single power law in this flux \cite{combined_fit_icrc}. For the tests performed here it is still modeled as a single power law with the parameters found in \cite{combined_fit_icrc}. In the final analysis this will be changed to a model with more degrees of freedom, for example a broken power law. \\
Most of the neutrinos in the event selections are neutrinos from cosmic-ray air showers. These are modeled using MCEq \cite{MCEq}. As a baseline, the H4a model for the cosmic-ray flux and Sibyll 2.3c for the hadronic interactions are chosen. The fluxes of conventional and prompt neutrinos are included separately, each with a free normalization. The uncertainty on the primary cosmic-ray flux is taken into account by allowing linear interpolation between the H4a and GST models. An additional parameter can change the spectral index of the cosmic-ray flux. The scheme developed by Barr et. al in \cite{Barr} is used to incorporate uncertainties on the hadronic interactions. A detailed description of the modeling of the atmospheric neutrino flux can be found in \cite{9.5yearsDiffuse}.\\
Finally there are some atmospheric muons which survive to the final level of the event selection. These are modeled with a dedicated simulation.

\subsection{Signal Modeling}
So far, there are six models included in the analysis. Three of them were already tested in \cite{GP_paper}, namely the \fermi{} model \cite{FermiLAT} with a spectral index of 2.72 and the KRA$_\gamma$ models with an exponential cutoff in primary cosmic-ray energy at 5 or \unit[50]{PeV} \cite{Gaggero_2015_2}. Additionally, we include the CRINGE model \cite{cringe} and two analytical models proposed by Fang and Murase \cite{FangMurase} where the spatial dependence and the energy dependence are separated. For the spatial model they once assume a constant product of gas and cosmic-ray density in the Galactic disc (FM-const) and once they assume that this product follows the distribution of supernova remnants. For the energy dependence of both models we assume a single power law with a spectral index of 2.7. The absolute normalization of each model is a free parameter in the fit.

\section{Results}
\begin{figure}[t!]
\centering
\includesvg[width=0.6\linewidth]{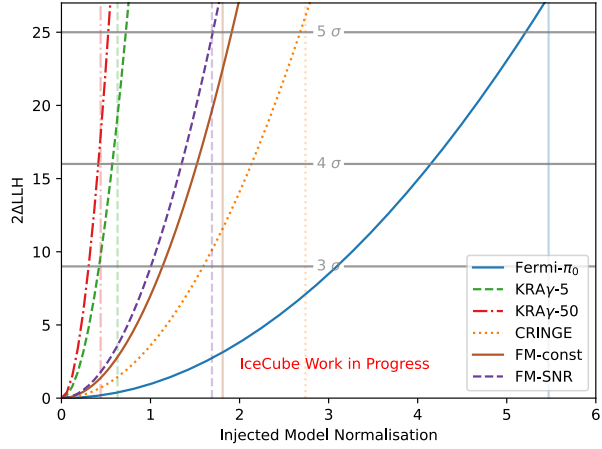}
\caption{Discovery Potentials for the six different models. Each model is injected at different normalizations and the likelihood difference between a free fit and a fit without Galactic contribution is calculated. The likelihood values corresponding to a 3, 4 and \unit[5]{$\sigma$} significance for a Galactic contribution are indicated by the gray lines. As the absolute normalizations of the models are quite different, the vertical lines show the norm which is fitted with this model, if the \fermi{} measurement from \cite{GP_paper} is injected.}
\label{dp}
\end{figure}

At first, the discovery potentials for each of the models are calculated. For this purpose the model is injected into the fit at different absolute normalizations and the expected significance for this model is calculated. The results can be found in \cref{dp,tab:dp}. As the absolute flux predictions of the models is quite different we chose some baseline normalizations for the models. For this purpose, the \fermi{} model is injected into the fit at the normalization found in \cite{GP_paper}. Each model is then separately fitted to this injection and the resulting best-fit normalizations are chosen as baselines. The baselines and the corresponding fluxes can also be found in \cref{tab:dp}. In \cref{dp} one can see these baselines as vertical lines. It can be seen, that these normalizations would all lead to significances above \unit[4]{$\sigma$} and some above \unit[5]{$\sigma$}.

\begin{table}[b!]
\centering
\begin{tabular}{ccccccc}
\hline
Model  &	\fermi{} & \krafive{} & \krafifty{} & CRINGE & FM-const & FM-SNR\\
\hline
\unit[3]{$\sigma$} DP & 3.09 & 0.42 & 0.31 & 1.59 & 1.14 & 1.00\\
\unit[4]{$\sigma$} DP & 4.15 & 0.57 & 0.41 & 2.13 & 1.52 & 1.35\\
\unit[5]{$\sigma$} DP & 5.21 & 0.72 & 0.53 & 2.68 & 1.91 & 1.70\\
\hline
baseline normalization & 5.47 & 0.63 & 0.44 & 2.74 & 1.81 & 1.69\\
\\
\begin{tabular}{@{}c@{}}$\phi_{@\unit[100]{TeV}}$ \\ in $10^{-18}$ GeV$^{-1}$ cm$^{-2}$ s$^{-1}$\end{tabular}
& 2.18 & 1.58 & 1.67 & 1.29 & 2.28 & 2.13\\
\hline

\end{tabular}
\caption{Discovery Potentials for the different models for the Galactic diffuse neutrino flux. The values are given as multiples of the predicted model flux. Additionally, the conversion to physical flux units is given for the chosen baseline normalizations. The given fluxes are all-sky integrated, per-flavor fluxes at \unit[100]{TeV}.}\label{tab:dp}
\end{table}

\begin{figure}[t!]
\centering
\includesvg[width=0.6\linewidth]{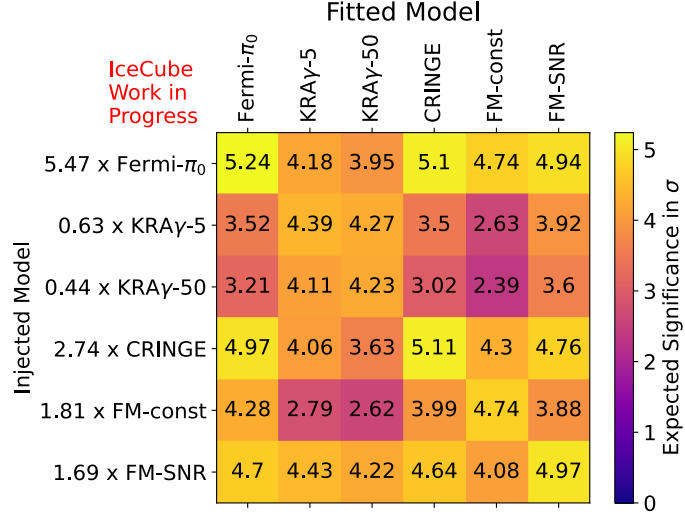}
\caption{Expected significance for a Galactic contribution if the model on the left is injected and the model at the top is fitted. The injected normalizations are chosen as explained in the text.}\label{sign}
\end{figure}

As the true model is unlikely to fully agree with one of the models, we perform an additional test which allows us to estimate the expected significance in case of a mismodeling. To achieve this, each model is injected at the baseline normalization. Then, a fit is performed in which another model is chosen and the expected significance for this model compared to no galactic contribution is calculated. The results can be found in the correlation matrix in \cref{sign}. It can be seen, that some models are quite similar (e.g. \fermi{} and CRINGE) and are expected to produce very similar significances. This shows that a slight derivation of the truth from the models is not affecting the performance of the analysis. But there are also some models like FM-const which, if they would be the only tested model, produce only a small significance in many cases. Therefore, it is important to test a wide range of different models to cover all possibilities.

\begin{figure}[t!]
\centering
\includesvg[width=0.6\linewidth]{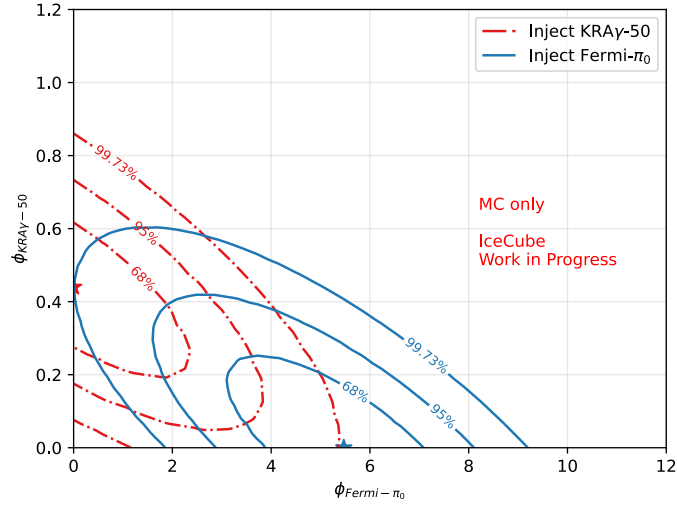}
\caption{Expected uncertainties for the model normalization of the \fermi{} and the \krafifty{} model. One of the models is injected and both models are included in the fit. Shown are the resulting 68, 95 and 99.73\% contours. }\label{llhscan}
\end{figure}

\begin{figure}[t!]
\centering
\includesvg[width=0.6\linewidth]{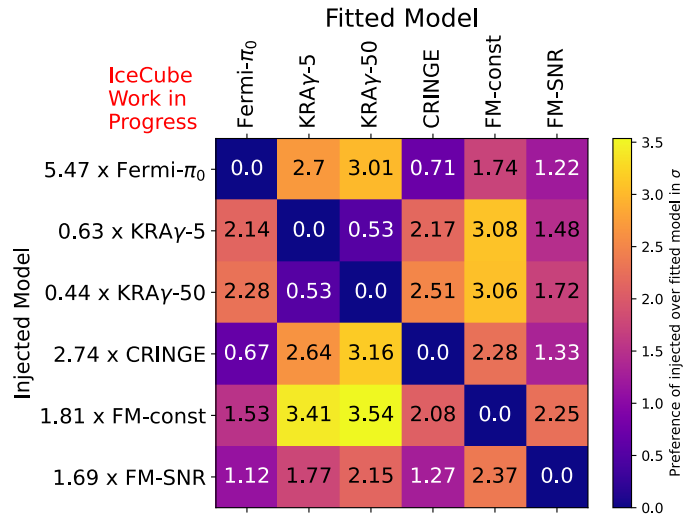}
\caption{Model discrimination power. For each injected model on the left the likelihood difference between a fit using this model and a fit using the model on the top is shown. The numbers show the expected preference of the injected model over the fitted model if the injected model would be the truth.}\label{modeldiscr}
\end{figure}

The difference in expected significances already gives a hint that some models predict different enough fluxes to allow some discrimination between the models. To test this in more detail, we inject one model and fit this model and another model to it. The result for \fermi{} and \krafifty{} can be seen in \cref{llhscan}. If the truth looks like \fermi{} (\krafifty{}), this model is preferred at \unit[3]{$\sigma$} (\unit[2.3]{$\sigma$}). The results for all combinations of models are summarized in \cref{modeldiscr}. Again this shows, that some models are very similar and we do not have the possibility to differentiate between them. Examples are the two KRA$_\gamma$ models or \fermi{} and CRINGE. We expect, that we can discriminate between these two groups with a significance of \unit[2-3]{$\sigma$}. The strongest discrimination can be found between the most extreme models KRA$_\gamma$ and FM-const. Here we expect to reject one of the models with more than \unit[3]{$\sigma$} if the other model is true.

\section{Conclusion}
We presented an analysis aimed to better understand the nature of the Galactic neutrino flux. For this we combine two independent event selections. We showed that the analysis is expected to measure this flux on a \unit[4]{$\sigma$}-level with a completely different approach then \cite{GP_paper}. We expect that the analysis can exclude some of the existing models with a significance of \unit[3]{$\sigma$}.

There are still some open issues. The cascade selection used in \cite{GP_paper} showed disagreements between data and simulation. This is not a problem for the analysis in [1], as it relies on scrambled data for background estimation. However, a forward-folding analysis as presented here, is more dependent on accurate simulations. Therefore, a modified version of this selection is under development and more details can be found in \cite{Rechav:2025icrc}. A second improvement to increase the agreement between data and simulation will be the usage of a more recent simulation using the most up to date model of the ice at the South Pole. Finally, for the study presented here, the effect of atmospheric self-veto \cite{NuVeto} is not taken into account. This effect decreases the number of atmospheric neutrinos ending up in the final level of the cascade selection, as they could be accompanied by muons from the same air shower.\\
As it was shown, the analysis has the power to show a preference for one model over others, if this model happens to be true. Therefore, it is interesting to test models which make some more extreme assumptions like the FM-const model, to limit the allowed parameter space. In addition to the FM-const model, which assumes a constant product of gas and cosmic-ray density, we plan to test the assumption of a constant cosmic-ray density by using a map of the gas distribution in the Galaxy as a template. \cite{Doerner:2025icrc} showed that considering anisotropic diffusion has a strong influence on the expected spatial distribution of Galactic diffuse emission. Based on this, we plan to test for the preferred anisotropy of cosmic ray diffusion. This will provide more information to understand the diffusion of cosmic rays in the Milky Way.


\bibliographystyle{ICRC}
\bibliography{references}

\clearpage

\section*{Full Author List: IceCube Collaboration}

\scriptsize
\noindent
R. Abbasi$^{16}$,
M. Ackermann$^{63}$,
J. Adams$^{17}$,
S. K. Agarwalla$^{39,\: {\rm a}}$,
J. A. Aguilar$^{10}$,
M. Ahlers$^{21}$,
J.M. Alameddine$^{22}$,
S. Ali$^{35}$,
N. M. Amin$^{43}$,
K. Andeen$^{41}$
% ...
\\
\\
% ...
$^{10}$ Universit{\'e} Libre de Bruxelles, Science Faculty CP230, B-1050 Brussels, Belgium \\
% ...
$^{16}$ Department of Physics, Loyola University Chicago, Chicago, IL 60660, USA \\
$^{17}$ Dept. of Physics and Astronomy, University of Canterbury, Private Bag 4800, Christchurch, New Zealand \\
% ...
$^{21}$ Niels Bohr Institute, University of Copenhagen, DK-2100 Copenhagen, Denmark \\
$^{22}$ Dept. of Physics, TU Dortmund University, D-44221 Dortmund, Germany \\
% ...
$^{35}$ Dept. of Physics and Astronomy, University of Kansas, Lawrence, KS 66045, USA \\
% ...
$^{39}$ Dept. of Physics and Wisconsin IceCube Particle Astrophysics Center, University of Wisconsin{\textemdash}Madison, Madison, WI 53706, USA \\
% ...
$^{41}$ Department of Physics, Marquette University, Milwaukee, WI 53201, USA \\
% ...
$^{43}$ Bartol Research Institute and Dept. of Physics and Astronomy, University of Delaware, Newark, DE 19716, USA \\
% ...
$^{63}$ Deutsches Elektronen-Synchrotron DESY, Platanenallee 6, D-15738 Zeuthen, Germany \\
$^{\rm a}$ also at Institute of Physics, Sachivalaya Marg, Sainik School Post, Bhubaneswar 751005, India \\
% ...

\subsection*{Acknowledgments}

\noindent
The authors gratefully acknowledge the support from the following agencies and institutions:
USA {\textendash} U.S. National Science Foundation-Office of Polar Programs,
U.S. National Science Foundation-Physics Division,
U.S. National Science Foundation-EPSCoR,
U.S. National Science Foundation-Office of Advanced Cyberinfrastructure,
Wisconsin Alumni Research Foundation,
Center for High Throughput Computing (CHTC) at the University of Wisconsin{\textendash}Madison,
Open Science Grid (OSG),
Partnership to Advance Throughput Computing (PATh),
Advanced Cyberinfrastructure Coordination Ecosystem: Services {\&} Support (ACCESS),
Frontera and Ranch computing project at the Texas Advanced Computing Center,
U.S. Department of Energy-National Energy Research Scientific Computing Center,
Particle astrophysics research computing center at the University of Maryland,
Institute for Cyber-Enabled Research at Michigan State University,
Astroparticle physics computational facility at Marquette University,
NVIDIA Corporation,
and Google Cloud Platform;
Belgium {\textendash} Funds for Scientific Research (FRS-FNRS and FWO),
FWO Odysseus and Big Science programmes,
and Belgian Federal Science Policy Office (Belspo);
Germany {\textendash} Bundesministerium f{\"u}r Forschung, Technologie und Raumfahrt (BMFTR),
Deutsche Forschungsgemeinschaft (DFG),
Helmholtz Alliance for Astroparticle Physics (HAP),
Initiative and Networking Fund of the Helmholtz Association,
Deutsches Elektronen Synchrotron (DESY),
and High Performance Computing cluster of the RWTH Aachen;
Sweden {\textendash} Swedish Research Council,
Swedish Polar Research Secretariat,
Swedish National Infrastructure for Computing (SNIC),
and Knut and Alice Wallenberg Foundation;
European Union {\textendash} EGI Advanced Computing for research;
Australia {\textendash} Australian Research Council;
Canada {\textendash} Natural Sciences and Engineering Research Council of Canada,
Calcul Qu{\'e}bec, Compute Ontario, Canada Foundation for Innovation, WestGrid, and Digital Research Alliance of Canada;
Denmark {\textendash} Villum Fonden, Carlsberg Foundation, and European Commission;
New Zealand {\textendash} Marsden Fund;
Japan {\textendash} Japan Society for Promotion of Science (JSPS)
and Institute for Global Prominent Research (IGPR) of Chiba University;
Korea {\textendash} National Research Foundation of Korea (NRF);
Switzerland {\textendash} Swiss National Science Foundation (SNSF).
JH, JBT, and WR acknowledge support from the Deutsche Forschungsgemeinschaft (DFG), via the Collaborative Research Center SFB1491 ”Cosmic Interacting Matters - From Source to Signal”.

\end{document}